\documentclass[aps,prl,twocolumn,superscriptaddress,showpacs]{revtex4}
\usepackage{amssymb}
\usepackage{amsmath}
\usepackage{graphicx}

\begin{document}

\title{Dirac and Normal Fermions in Graphite and Graphene: \\
Implications to the Quantum Hall Effect}
\author{Igor A. Luk'yanchuk}
\affiliation{University of Picardie Jules Verne, Laboratory of Condensed Matter Physics,
Amiens, 80039, France}
\affiliation{L. D. Landau Institute for Theoretical Physics, Moscow, Russia}
\author{Yakov Kopelevich}
\affiliation{Instituto de F\'{\i}sica "Gleb Wataghin", Universidade Estadual de Campinas,
Unicamp 13083-970, Campinas, Sao Paulo, Brazil}
\date{\today}

\begin{abstract}
Spectral analysis of the Shubnikov de Haas (SdH) magnetoresistance oscillations
 and of the Quantum Hall Effect (QHE) measured in quasi-2D
highly oriented pyrolytic graphite (HOPG) [Phys. Rev. Lett. 90, 156402
(2003)] reveals two types of carriers: normal (massive) electrons with Berry
phase $0$ and Dirac-like (massless) holes with Berry phase $\pi$. We
demonstrate that recently reported integer- and semi-integer QHE for
bi-layer and single-layer graphenes take place simultaneously in HOPG
samples.
\end{abstract}

\maketitle

Observations of the quantum Hall effect (QHE) and magnetic-field-driven
metal-insulator transition in quasi-2D highly oriented pyrolytic graphite
(HOPG) \cite{Kopelevich_2003,Kopelevich_1999} indicate that much in the physics
of graphite had been missed in the past and had triggered a considerable
interest in graphite once again. Analyzing the quantum de Haas van Alphen
(dHvA) and Shubnikov de Haas (SdH) oscillations in bulk HOPG samples we
experimentally proved \cite{Lukyanchuk_2004} that, besides the conventional
electronic charge carriers with the massive spectrum $E=p^2/2m$,
massless (2+1)-dimensional fermions (holes) with Dirac-like linear spectrum $%
E=\pm v|p|$ and non-trivial Berry phase $\pi$ do exist in graphite.
It is likely that these Dirac-type carriers are responsible for the
strongly correlated phenomena predicted by theory
\cite{Gonzalez96,Khveshchenko01,Gorbar02}. The results of
\cite{Lukyanchuk_2004} were confirmed by the recent angle resolved
photoemission spectroscopy (ARPES) experiments performed on HOPG
\cite{Zhou06,Zhou06N}, that provide an unambiguous experimental
evidence for coexistence of Dirac-like holes and normal (massive)
electrons located at H and K poins of the Brillouin zone,
respectively.

\begin{figure}[b!]
\centering
\includegraphics [width=7cm] {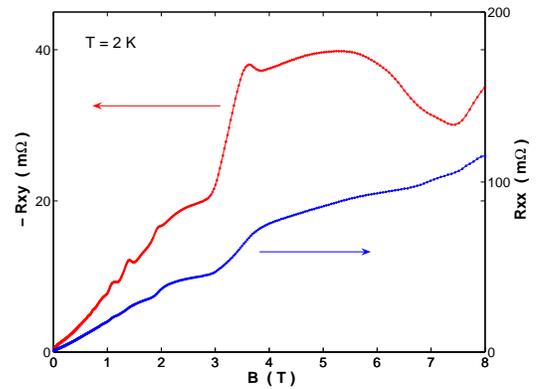}
\caption{Field dependence of basal-plane magnetoresistance $R_{xx}(B)$ and
Hall resistance $R_{xy}(B)$ measured for the HOPG-3 sample with magnetic
field B applied along the sample c-axis [1]. }
\label{FigHOPG}
\end{figure}

Very recently,  remarkable progress in the technology of ultrathin films
consisting of one \cite{Novoselov_2005,Zhang_2005}, two \cite{Novoselov_2006}
or several \cite{Morozov_2005} graphite monolayers (graphenes) was achieved.
Analysis of unconventional half-integer QHE and Berry phase $\pi$ of SdH
oscillations in graphene led to the conclusion that charge carriers in graphene
have the nature of Dirac-like massless fermions \cite%
{Novoselov_2005,Zhang_2005}. At the same time, bilayer graphene
demonstrates a new type of the integer QHE in which the last
(zero-level) plateau is missing, and therefore the charge carriers
are described by the chiral fermions having the Berry phase $2\pi$
and the energy spectrum is represented by two touching parabolas
$E(p)= \pm p^2/2m$ \cite{Novoselov_2006}.

\begin{figure}[!b]
\centering
\includegraphics [width=7cm] {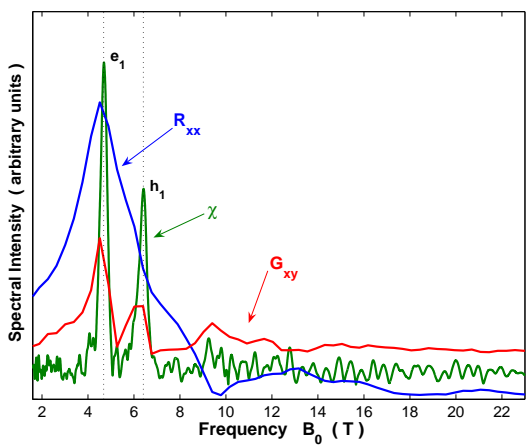}
\caption{Spectral intensity of SdH oscillations of magnetoresistance $%
|R_{xx}(B_0)|$ and of Hall conductance $|G_{xy}(B_0)|$ in HOPG-3 sample in
comparison with dHvA oscillations of susceptibility $|\protect\chi(B_0)|$ in
HOPG-UC sample \protect\cite{Lukyanchuk_2004}. Peaks $e_1$ and $h_1$
correspond to the normal electrons and Dirac holes, respectively}
\label{FigSpectr}
\end{figure}

In the present work we apply our filtering and phase analysis
procedures \cite{Lukyanchuk_2004} to analyze the SdH oscillations and
the QHE effect in HOPG aiming to clarify how these effects correlate
with those studied in single and bilayer graphene systems. In
particular, we demonstrate that normal and Dirac-like fermions seen
in the bulk graphite are responsible respectively for the integer
and semi-integer precursors of the QHE. To check  the independence
of these conclusions on the windowing and frequency filtration
cut-off we used an alternative, insensitive to filtration procedure
method of two-dimensional phase-frequency  analysis, described in
detail in \cite{Lukyanchuk_2004}.

We analyze the  field dependencies of the basal-plane
resistance $R_{xx}(B)$ and the Hall resistance $R_{xy}(B)$ reported in Ref.
\cite{Kopelevich_2003} for the sample labeled as HOPG-3 (noting that the HOPG-UC
sample, similar to HOPG-3, has been studied in Ref. \cite{Lukyanchuk_2004}).

As seen from Fig. \ref{FigHOPG}, $R_{xy}(B)$ demonstrates several QHE-like
plateaus at high fields, also observed in various HOPG samples by other
groups \cite{Novoselov_2004,Niimi05}. However, unlike the conventional QHE, $%
R_{xx}(B)$ does not drop to zero at the positions of these plateaus. Only
weak SdH oscillations $\Delta R_{xx}(B)$ [3] are seen superimposed on almost
linear $R_{xx}(B)$ dependence. This can be explained either by imperfection
of the QHE itself or by the existence of a background resistance of unknown origin.
It is interesting to note that QHE with  similar behavior of $R_{xx}(B)$
has been reported for Bechgaard salts \cite{Hannahs89,Poilblanc87} and for
(Bi$_{1-x}$Sb$_x$)$_2$Te$_3$ layered semiconductors \cite{Elefant98,Sasaki01}.

Because quantum oscillations are periodic functions of the inverse field $%
B^{-1}$, the inverse-field spectral analysis is the appropriate tool to
discriminate between different groups of carriers involved in
oscillations. Fig. \ref{FigSpectr} presents the spectral intensity of SdH
oscillations of the resistance $R_{xx}(B^{-1})$ and the Hall conductance $%
G_{xy}(B^{-1}) \equiv R_{xy}^{-1}(B^{-1})$, (in both cases polynomial
background has been subtracted) together with analyzed in \cite%
{Lukyanchuk_2004} spectral intensity of dHvA oscillations of magnetic
susceptibility $\chi(B^{-1})$, ($\chi = dM/dB$).

\begin{figure}[!b]
\centering
\includegraphics [width=7.5cm] {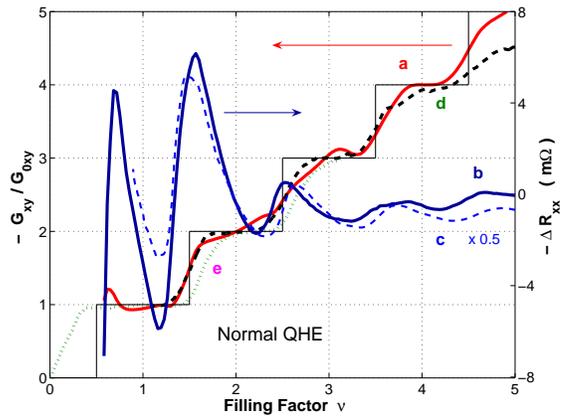}
\caption{Hall Conductance $G_{xy}$ (solid line a) and
$\Delta\protect\sigma_{xx} \sim - \Delta R_{xx}$ (solid line b) for
HOPG-3 sample as a function of the normalized "filling factor"
$\protect\nu=B_0/B$ (were $B_0=4.68T$ is the SdH oscillation
frequency). Normal electrons give the dominant contribution in the
quantum oscillations. The profile of $G_{xy}(\protect\nu)$ is fitted
by the integer QHE staircase; the minima of $\Delta
\protect\sigma_{xx}$
correspond to the QHE plateaus. Hall conductance is normalized on $%
G_{0xy}=28\Omega^{-1}$ equal to the step between QHE plateaus. The
results are compared with SdH oscillations in thick film of graphite
(dashed line c) \protect\cite{Novoselov_2004}(having approximately
the same $B_0=4.65T$) and with the QHE in bilayer graphene
\protect\cite{Novoselov_2006} obtained as a function of both
magnetic field (dashed line d) ($\protect\nu=B_0/B$, $B_0=27.8T$)
and
concentration (dotted line e) ($\protect\nu=n/n_0$ with $n_0=1.7\cdot10^{12}cm^{-2}$ at $%
B=20T$)}
\label{FigNormal}
\end{figure}

Fig. \ref{FigSpectr} shows that two peaks in the susceptibility spectrum
observed at 4.68T and at 6.41T are also resolved in the spectrum of $%
G_{xy}(B^{-1})$. The higher-frequency magnetoresistance peak
manifests itself as the right-shoulder structure of $R_{xx}(B^{-1})$
spectrum. From  comparative analysis
of dHvA and SdH oscillations in \cite{Lukyanchuk_2004} we identified these peaks as
originating from normal (electrons) and Dirac (holes) carriers.
Here, we re-affirm this conclusion in a different way, namely by
means of the comparative analysis of longitudinal and Hall
conductance oscillations, which is based on the separate study of the
contributions from both types of carriers.

\begin{figure}[!b]
\centering
\includegraphics [width=7.5cm] {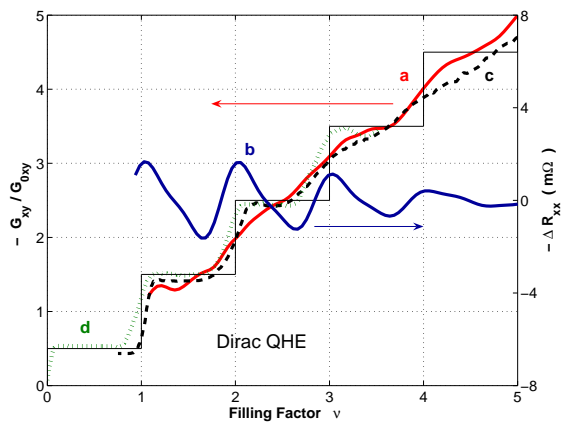}
\caption{Solid lines a and b : the same as in Fig. \protect\ref{FigNormal} (with $%
B_0=6.41T$, $G_{0xy}=19\Omega^{-1}$) but after filtering elimination
of the normal electrons contribution. The $G_{xy}(\protect\nu)$
profile and positions of the minima in $\Delta \protect\sigma_{xx}
\sim - \Delta R_{xx}$ are coherent with the Dirac QHE staircase. The
results are compared with the Dirac QHE in graphene monolayer
obtained both as a function of magnetic field ($\protect\nu=B_0/B$
with $B_0=6.67T$) (dashed line c) \protect\cite{Zhang_2005}
and carrier concentration ($\protect\nu=n/n_0$ with $n_0=1.5%
\cdot10^{12}cm^{-2}$ at $B=14T$)(dotted line d)
\protect\cite{Novoselov_2005}} \label{FigDirac}
\end{figure}

Fig. \ref{FigNormal} (curves a and b) shows the raw data of $G_{xy}(B^{-1})$
and $\Delta R_{xx}(B^{-1})$ (after background signal subtraction) in which
the normal electron contribution dominates in the
magnetotransport quantum
oscillations \cite{Lukyanchuk_2004}. Extraction of the Dirac-like carriers
signal from these data was the most challenging task that we performed by
stop-band filtering of $G_{xy}(B^{-1})$, eliminating the frequency windows
related to the electron spectral peak and its harmonics (see Fig. \ref{FigSpectr}%
) and using the pass-band filtering of $\Delta R_{xx}(B^{-1})$ with the
corresponding hole frequency windowing. The resulting contributions from
Dirac-like carriers are presented in Fig. \ref{FigDirac} (curves a and b).

Note that the sign of $\Delta R_{xx}$ is reversed in order to present the
oscillating part of the conductivity $\Delta\sigma_{xx}$, because in our
case of $\rho_{xx} \gg \rho_{xy}$ and  $\Delta \sigma_{xx}$ is related to $%
\Delta R_{xx}$ as:
\begin{equation}
\Delta \sigma_{xx}=\Delta{\frac{\rho_{xx} }{\rho_{xx}^2 +
\rho_{xy}^2}} \simeq - {\frac{\Delta\rho_{xx} }{\rho_{xx}^2}}\sim -
\Delta R_{xx}. \label{SdH}
\end{equation}

The situation is just opposite to the conventional QHE case with
giant oscillations of $R_{xx}(B)$ when $\rho_{xx} \ll \rho_{xy}$ and
$\sigma_{xx}$ behaves in the same way as $R_{xx}$. Note also that
the amplitude of SdH oscillations associated with Dirac-like
fermions (Fig. \ref{FigDirac}, curve b) is three times smaller than
that for normal carriers (Fig. \ref{FigNormal}, curve b).

To facilitate the comparative analysis of the results obtained for various
graphite (graphene) samples, we present all the data as a function of the
normalized filling factor $\nu = B_0/B$, where $B_0$ corresponds to the SdH
oscillation frequency. The conductance $G_{xy}$ is normalized by
$G_{0xy}$, ($\simeq 28\Omega^{-1}$ for electrons and $19\Omega^{-1}$
for holes) corresponding to the step between subsequent QHE plateaus.

Fig. \ref{FigNormal} (curve c) includes SdH oscillations reported
for an $8\mu m $ thick HOPG sample taken from the inset of Fig. 4 in
\cite{Novoselov_2004} after the polynomial background signal
subtraction. The coincidence of the SdH oscillation frequency, phase
and the asymmetric form of the signal with those measured in our
HOPG sample suggests that the same Fermi surface electron pocket is
responsible for SdH oscillations in both samples.

Before we proceed with the data analysis, we note that Landau Level
(LL) quantization spectra for normal and Dirac-like carriers are
different. In the normal carrier case the equidistant LLs
$E_n=(e\hbar /m_\perp c)B(n+1/2)$ \cite{Landau3} are separated by
the gap $E_0=e\hbar B/2m_\perp c$ from $E=0$
whereas in the Dirac-like case $E_n=\pm v \sqrt{2e\hbar B n/c}$ \cite%
{Landau4}, and the Lowest Landau Level (LLL) is located exactly at $E_0=0$.
This leads to two important experimental consequences.

(i) SdH oscillations of conductivity \cite{Lukyanchuk_2004}
\begin{equation}
\Delta \sigma_{xx}(B) \simeq - A(B) \cos[2\pi({\frac{B_0
}{B}}-\gamma+\delta) ]
\end{equation}
(where $A(B)$ is the non-oscillating amplitude) acquire the phase factor
either $\gamma = 1/2$ or $\gamma= 0$ for normal and Dirac carriers,
respectively. The additional phase factor $\delta$ governed by the curvature
of Fermi surface is quite small: $|\delta|<1/8$ (in 2D $\delta=0$) and can
be neglected.

(ii) QHE plateaus occur either in integer (normal carriers) or in
semi-integer (Dirac fermions) way and can be expressed via the phase factor $%
\gamma$ as \cite{Gusynin05,Peres06}:
\begin{equation}
G_{xy}=\mu g_s{\frac{e^2}{h}}(n+{\frac{1 }{2}}-\gamma)  \label{Hall}
\end{equation}
where $\mu=\mp 1$ for electrons/holes, and $g_s$ is the (iso)-spin
degeneracy factor.

\begin{figure}[!b]
\centering
\includegraphics [width=7cm] {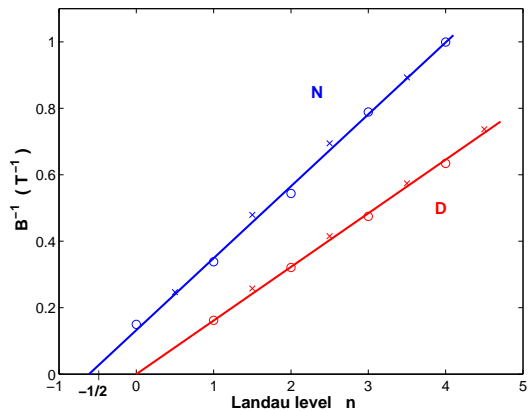}
\caption{Values of $B^{-1}$ for the normal (N) and Dirac-like (D)charge
carriers at which n$th$ Landau level crosses the Fermi level. These values
are found as maxima (o) in the SdH oscillations of magnetoconductivity $%
\Delta \protect\sigma_{xx}(B)$. The minima positions (x) in $\Delta \protect%
\sigma_{xx}(B)$ are shifted by $1/2$ to the left. Linear extrapolation of
the data to $B^{-1} = 0$ (solid lines) allows for the phase factor ($-
\protect\gamma + \protect\delta$, Eq. (2)) definition: $\protect\gamma=1/2$,
$\protect\delta\simeq-1/8$ for normal electrons and $\protect\gamma=0$, $%
\protect\delta=0$ for Dirac-like holes.}
\label{FigFan}
\end{figure}

The phase factor $\gamma$ is defined for an arbitrary spectrum
$E(p)$ by the quasiclassical quantization condition of the Fermi
surface cross-section $S(E_f)=(n+\gamma)2\pi{eB/\hbar c}$ ($n\gg
1$). This factor is uniquely related to the topological Berry phase
$\Phi_B=k\pi$ acquired by a fermion, moving around $S(E_f)$
\cite{Mikitik99}: $\gamma$ equals either $1/2$ for even $k$ (normal
carriers) or $0$ for odd $k$ (Dirac-like fermions). For example, the
two-parabola spectrum of bilayer graphite $E(p)= \pm p^2/2m_{\perp}$
is quantized in magnetic field as: $E_n=\pm(e\hbar /m_\perp
c)B\sqrt{n(n-1)}$ \cite{McCann_2006}. This gives an unconventional
doubly degenerate LLL with $E_0=E_1=0$ but for higher $n$ the system
recovers the behavior of normal carriers having $\gamma=1/2$ (with
$n\rightarrow n-1 $ in (\ref{Hall})).

We obtain the phase factors $\gamma$ and $\delta$ by plotting the
inverse field values at conductivity oscillation maxima (minima) as
a function of their number (number - 1/2), as shown in Fig.
\ref{FigFan}. The linear extrapolation of data points to $B^{-1}=0$
unambiguously determines the
phase factors: $\gamma=1/2$, $\delta\simeq-1/8$ for data depicted in Fig. %
\ref{FigNormal} (curve b) and $\gamma=0$, $\delta=0$ for data shown in Fig. %
\ref{FigDirac} (curve b). Thus, we identify here normal (3D) and
Dirac-like (2D) charge carriers in agreement with our previous
results obtained for HOPG-UC sample \cite{Lukyanchuk_2004}.

Curves (a) in Figs. \ref{FigNormal} and \ref{FigDirac} provide
evidence for the staircase behavior (quantization) of $G_{xy}$ for
both normal and Dirac-like fermions. As expected for QHE regime, the
minima in longitudinal conductivity $\sigma_{xx}$ coincide with the
plateau positions in $G_{xy}$.

In the case of normal carriers , Fig. \ref{FigNormal} (curve a), the
quantization of $G_{xy}$ corresponds to the integer QHE, as follows
from Eq. (\ref{Hall}) for $\gamma=1/2$. In the same Fig.
\ref{FigNormal} we presented the QHE staircases of $\sigma_{xy}$,
reported in \cite{Novoselov_2006} for
bilayer graphite as function of both inverse field $B^{-1}$ (Fig. \ref%
{FigNormal}, curve d) and carrier concentration $n$ (Fig.
\ref{FigNormal}, curve e). The scales of $B^{-1}$ and $n$ were
normalized to the corresponding oscillation periods of
$\sigma_{xx}$. The Hall conductivity was normalized to the step
between subsequent plateaus in $\sigma_{xy}(B)$.
The comparative analysis indicates that all three dependencies, i. e. $%
G_{xy}(B^{-1})$ for HOPG, $\sigma_{xy}(B^{-1})$ and $\sigma_{xy}(n)$
for bilayer graphene are equally well fit to the integer QHE theory
with $\gamma = 1/2$. We stress, however, that our data don't allow
us to distinguish between the conventional integer QHE with Berry
phase $0$ and the chiral one with Berry phase $2\pi$, proposed for
bilayer graphite \cite{Novoselov_2006}. This is because both models
imply $\gamma=1/2$ and the same integer QHE staircase at LL $n>1$.
The only difference between them is the absence of zero level
plateau for chiral fermions whose (non)existence we cannot verify at
the moment.

It is interesting to note that the absolute values of Hall steps
$\Delta \rho _{xy}^{-1}$, estimated (similar to
\cite{Kopelevich_2003}) as $(10\, \mathrm{k}\Omega /\square )^{-1}$
for electrons and $(15\, \mathrm{k}\Omega /\square )^{-1}$ for holes
are approximately equal to the two-spin degenerate QHE step $2e^2/h
\simeq (12.9 \thinspace \mathrm{k}\Omega /\square)^{-1}$ ($g_s=2$ in
(\ref{Hall})) and are twice as small as the two-spin two-valley
degenerate QHE step with $g_s=4$, reported for graphene
\cite{Zhang_2005,Novoselov_2005} and bilayer graphite
\cite{Novoselov_2006}.

To conclude, the results reported in this letter give  new insight
on the behavior of Dirac and normal fermions in graphite. Namely, we
demonstrate the coexistence of QHE precursors associated with normal
electrons and Dirac-like holes. To the best of our knowledge this is
the first observation of the simultaneous occurrence of integer and
semi-integer quantum Hall effects.

This work was supported by Brazilian scientific agencies FAPESP,
CNPq, CAPES, and French agency COFECUB. We thank to Prof. M. G.
Karkut for the useful discussions.

\end{document}